\documentclass[adraft,creativecommons,noreriv]{eptcs}
\providecommand{\event}{ML 2017} % Name of the event you are submitting to

\usepackage{hyphenat}
\usepackage{mdwtab}
\usepackage{mdwlist}
\usepackage{amssymb}
\usepackage{amsmath}
\usepackage{mathenv}
\usepackage{comment}
\usepackage{url}
\usepackage{listings}

\lstdefinestyle{sOcaml}{language=[Objective]Caml,
  literate={+}{{$+\:$}}1 {/}{{$/$}}1 % { * }{{$*$}}1
           {=}{{${=}$}}1
           {>}{{$>$}}1 {<}{{$<$}}1
           {<>}{$\not=\ $}1
           {->}{{$\rightarrow$}}2 {>=}{{$\geq$}}2 {<-}{{$\leftarrow$}}2
           {<=}{{$\leq$}}2
           {==>}{{$\mapsto$}}2
           {===}{{$\equiv$}}1
           {|}{{$\mid$}}1
           {'a}{$\alpha$}1
           {+'a}{$\textrm{+}\alpha$}1
           {'b}{$\beta$}1
           {+'b}{$\textrm{+}\beta$}1
           {'c}{$\gamma$}1
           {'e}{$\epsilon$}1
           {'w}{$\omega$}1
           {'w.}{$\forall\omega.\ $}2
           {<--}{{$\longleftarrow$}}2
           {TRB}{\mbox{\ensuremath\lceil}}1
           {TRE}{\mbox{\ensuremath\rceil}}1
           {CCC}{\ensuremath{\mid\mid\mid}}1
           {...}{\ldots}2
           {\#\#\#}{{$\leadsto$}}3
}
\lstnewenvironment{code}{\lstset{basicstyle={\sffamily}}}{}

\lstMakeShortInline[columns=fullflexible]|

\newcommand{\oleg}[1]{{\it [Oleg says: #1]}}
\newcommand{\aside}[1]{\ignorespaces}

\newtheorem{exercise}{Exercise} 

\begin{document}

\title{Effects Without Monads: Non-determinism\\Back to the Meta Language}
\author{Oleg Kiselyov
\institute{Tohoku University, Japan}
\email{oleg@okmij.org}}
\def\authorrunning{Oleg Kiselyov}
\def\titlerunning{Effects Without Monads}
\maketitle

\begin{abstract}
We reflect on programming with complicated effects, recalling
an undeservingly forgotten alternative to monadic programming and
checking to see how well it can actually work in modern functional
languages.

We adopt and argue the position of factoring an effectful program into
a first-order effectful DSL with a rich, higher-order `macro'
system. Not all programs can be thus factored. Although the approach
is not general-purpose, it does admit interesting programs.  The
effectful DSL is likewise rather problem-specific and lacks
general-purpose monadic composition, or even functions. On the upside,
it expresses the problem elegantly, is simple to implement and reason
about, and lends itself to non-standard interpretations such as code
generation (compilation) and abstract interpretation. A specialized
DSL is liable to be frequently extended; the experience with the
tagless-final style of DSL embedding shown that the DSL evolution can
be made painless, with the maximum code reuse.

We illustrate the argument on a simple but representative example of a
rather complicated effect~-- non-determinism, including committed
choice. Unexpectedly, it turns out we can write
interesting non-deterministic programs in an ML-like language just as
naturally and elegantly as in the functional-logic language Curry~--
and not only run them but also statically analyze, optimize and compile.
The richness of the Meta Language does, in reality, compensate
for the simplicity of the effectful DSL.

The key idea goes back to the origins of ML as the Meta Language for
the Edinburgh LCF theorem prover.  Instead of using ML to build theorems,
we now build (DSL) programs.
\end{abstract}

\begin{comment}
Andres Loeh talk at Dagstuhl 2018: he has seen type signatures in client
code that had 40-50 lines of MonadXXX constraints.
``We need `meaningful interfaces' rather than `generic interfaces'''
``Don't make it generic until you have used it twice.''
\end{comment}

\section{Introduction}

How to cope with the complexity of writing programs? How to structure
computations? Many methodologies have been proposed over the decades:
procedures, structured programming, OOP, AOP, algebraic specifications
and modules, higher-order functions, laziness~-- and, lately, monads
and their many generalizations. Although monads are not the only way
to organize (effectful) computations, they are by all accounts
receiving disproportionate attention (just do a quick Google
search). In ML, monads have been
introduced more \cite{swamy11monad} or less
\cite{multi-stage-monad} formally and underlie the widely used OCaml
libraries Lwt\footnote{\url{http://ocsigen.org/lwt/}}
%% Tutorial on Lwt
%% \url{https://mirage.io/wiki/tutorial-lwt}}
and Async.\footnote{\url{https://github.com/janestreet/async}}

This position paper seeks to draw attention to a non-monadic
alternative: Rather than structuring (effectful) programs as
monads~-- or applicatives, arrows, etc.,~-- we approach programming as
a (micro) language design. We determine what data
structures and (effectful) operations seem indispensable for the problem at
hand~-- and design a no-frills language with just these domain-specific
features. We embed this bare-bone DSL into OCaml, relying on OCaml's
extensive facilities for abstraction and program composition (modules,
objects, higher-order functions), as well as on its parsing and type checking. 

We state the main points of our argument in \S\ref{s:argument}. We
hasten to say that the key insight is rather old
\cite{wand-specifications}, quite resembling algebraic specifications
(see Wirsing's comprehensive survey \cite{Wirsing-specifications}).
In fact, it is the insight behind the original ML, as a scripting
language for the Edinburgh LCF theorem prover \cite{ML-original}~--
only applied to programs rather than theorems. What has not been clear
is how simple an effectful DSL may be while remaining useful.  How
convenient is it, especially compared to the monadic encodings? How
viable is it to forsake the generality of first-class functions and
monads and what benefits may come? We report on an experiment set out
to explore these questions.

The present paper follows the style of  
Hughes \cite{hughes-why}, Hudak \cite{hudak-building} and
Goguen \cite{Goguen-noHO}~-- or, for mathematically inclined,
P{\'o}lya \cite{polya-how}~-- arguing from
examples. Just like those arguments, it is indeed hard to grasp the
limitations and applicability, and it is hard to formalize. Problem
solving in general is a skill to learn rather than an algorithm to
implement; it is inherently informal. Even in mathematics, how to
prove theorems is an art and a judgement; one acquires it not by following
rigorous descriptions but by reading existing proofs and doing
exercises. This is the format we follow in the paper.

\subsection{Motivation}
\label{s:motivation}

This present paper comes as the result of decade-long long
experience with the 
tagless-final style of DSL embedding \cite{tagless-final-oxford}
and the re-discovering\footnote{\relax
\url{http://okmij.org/ftp/Computation/having-effect.html}}
and polishing of extensible effects (\cite{exteff,freer}). It was prompted
however by the following message, posted on the Caml-list by 
Christoph H{\"o}ger in March 2017:\footnote{\relax
Christoph H{\"o}ger: Transforming side-effects to a monad.
Posted to \url{caml-list@inria.fr} on
Thu, 23 Mar 2017 20:56:16 +0100}
\begin{quotation}
``Assume a simple OCaml program with two primitives that can cause
side-effects:
\begin{code}
let counter = ref 0
let incr x = counter := !counter + x ; !counter
let put n = counter := n; !counter
put (5 + let f x = incr x in f 3)
\end{code}

This example can be transformed into a pure program using a counter
monad (using |ppx_monadic| syntax):
\begin{code}
do_;
  i <-- let f x = incr x in f 3 ;
  p <-- put (5 + i)
  return p
\end{code}
For a suitable definition of bind and return, both programs behave
equivalently. My question is: How can one automatically translate a
program of the former kind to the latter?''
\end{quotation}

The message left me puzzled about its author's goals and
motivations. It is hard to imagine he preferred the monadic program
for its verbose notation, replete with irrelevant names like |i| and
|p|. Was the author after purity? That is a mirage we lay bare in
\S\ref{s:objections}. Was he attracted to the separation of effectful
and non-effectful code and the possibility of multiple interpretations
of effects (`the overriding of the semicolon')?  These good properties
are not unique to monads. The other, often forgotten, ways of dealing
with side-effects ought to be more widely known.

The second cue came about a month later, observing students
solving an exercise to compute all permutations of a given list of
integers. The reader may want to try doing that in their
favorite language. Albeit a simple exercise, the code
is often rather messy and not \emph{obviously} correct.  In the
functional-logic language Curry \cite{CurryReport} built around
non-determinism, computing a permutation is strikingly elegant: 
mere |foldr insert []|. It is the re-statement of the specification: 
a permutation
is moving the elements of the source list one-by-one into \emph{some}
position in the initially empty list.  The code immediately tells that
the number of possible permutations (possible choices of permutations)
of |n| elements is |n!|. From its
very conception in the 1950s \cite{rabin-scott}, non-determinism
was called for to write clear specifications~-- and then to make them
executable.  Can we write the list permutation code just as elegantly
in a language that
was not designed with non-determinism in mind and offers no support
for it? How far can we extend it?

\subsection{Points to Argue}
\label{s:argument}

The primary goal for the paper is to report on the experiment set out to
explore the viability and consequences of a particular method of
writing effectful programs. Although all the ingredients have been
known (some of them for so long that they are almost forgotten), how
well the approach actually works for interesting problems 
can only be determined empirically.

Along with the describing the experiment and its ramifications, we also
offer an argument: why this approach is worth exploring in the
first place. The points of the argument, reverberating throughout the
paper, are collected below:

\begin{description}
\item[Effects are not married to monads]

The discussion after the first presentation of this paper, at the ML
Family workshop 2017, was one of many indications that monads have a
special, almost cult status in the minds of functional programmers. There
is no doubt that monads clearly delineate effectful computations, in
syntax and in types, and offer the reasoning principles (equational
laws) about effectful programs. What many do not seem to realize is
that these benefits are not unique to monads, or that not
all effects are expressible with monads,\footnote{The fact that not all
  effects are expressible as monads was noted already by 
Wadler \cite{Wadler-monad-cont}. That fact has motivated the development
of various monad-\textit{like} 
(relative monads \cite{relative-monads}, parameterized
monads \cite{atkey-parameterised}) and unlike interfaces
(applicatives \cite{Applicative} and arrows \cite{Arrows}).}
or that the
flexibility of the monadic encoding (`overriding semicolon') is
limited. Code generation and abstract interpretation, 
for example, do not fit the monadic framework (see \S\ref{s:no-do}).

Although this point may be obvious to some, the caml-list (as noted in 
\S\ref{s:motivation}), Reddit, Stack Overflow, etc. discussion places 
are awash with misunderstandings and irrational exuberance
towards monads.\footnote{\relax
Here is a small sample of links
\url{https://stackoverflow.com/questions/44965/what-is-a-monad}
\url{https://news.ycombinator.com/item?id=16422452}
\url{https://news.ycombinator.com/item?id=17645277}
\url{https://news.ycombinator.com/item?id=16419877}}
The argument pointing out their proper place and
limitations is worth repeating. 

\item[Separate rather than combine higher-order and effects]

The present paper is an exploration of a less common approach to
writing reusable, properly abstracted effectful programs. Rather than
combining effectful operations with 
modules, objects, higher-order functions, we separate them. First we
determine the data types and operations needed for the problem
at hand, and define the corresponding domain-specific language (DSL).
The language is often first-order, and its operations have a number of
\emph{domain-specific} effects (such as references to
application-specific context, communication, logging, etc). 

Since the DSL is intentionally without abstraction or syntactic
sugar~-- just enough to express the problem at hand, however
ungainly~-- programming in it directly is a chore. That is why we
endow it with a very expressive `preprocessor', by embedding into
a metalanguage with rich abstractions like functions,
definitions, modules, etc.

Not all problems can be thus factored into a first-order DSL and a
higher-order metalanguage (e.g., the factoring does not support
arbitrary higher-order effectful functions). Therefore,
how well the factoring works in practice and if it is worth
paying attention to become empirical questions. The paper describes
one case study, exploring how far we can push this approach.

\item[General vs. specific: it is a trade-off]

The approach to be evaluated in the present paper~-- the factoring of an
effectful program into a simple DSL and a rich preprocessor~-- is not
general purpose. The non-deterministic DSL is not general purpose
either: we introduce only those data types and operations that are needed
for the problem at hand and its close variations. Domain-specific,
narrow solutions should not be always looked down upon, we argue.

The compelling case for (embedded) DSLs has been already made, by 
Hudak \cite{hudak-building}. The present paper is another case study.
We also demonstrate, in \S\ref{s:no-do}, one more advantage of an
embedded DSL: the ability to evaluate the same DSL code in several ways.
We can not only (slowly) interpret the code, but also perform static
analyses such as abstract interpretation, and generate (faster)
code.

The advantages of the domain-specific approach have to be balanced
against the applicability (how wide is the domain, does it let us do
something interesting or practically significant) and extensibility
(how easy it is to extend the program and the domain and reuse the
existing code). This is a trade-off, which the case study in the present
paper is to help evaluate.

Thus we do not argue that the domain-specific approach is `better'. We
do argue, however, against the presumption (evidenced in the received
comments on the drafts of this paper) that one always has to strive for the
general solution. Premature generalization and abstraction, like
premature optimization, is not a virtue.

\item[Try domain-specific first]
When deciding which approach to effects fits the problem at
hand, we advocate trying a specialized solution (such as the DSL factoring)
first. Typically, whether it works out or not becomes clear very soon;
if it does not, little effort is wasted because the approach is so simple.
\end{description}

The structure of the paper is as
follows. \S\ref{s:list-perm} describes the main experiment: can we
write the list permutation in OCaml as elegantly as in Curry.
Specifically, \S\ref{s:NDet} introduces the intentionally
very simple, essentially first-order, DSL for the specific domain of
non-deterministic computations on integer lists, and \S\ref{s:Perm}
uses it to express the list permutation. The readers can see for
themselves how good (simple, understandable, close to Curry) it looks.
A way to make the DSL embedding
seamless is described in \S\ref{s:no-functor}.
Several `standard' implementations of the DSL are discussed
in \S\ref{s:list-impl}, whereas \S\ref{s:sorting} extends the DSL and its
implementation with the committed choice and presents another
classical Curry example: slow sort.  Thus the factoring of an effectful
computation into a first-order DSL and a powerful metalanguage turns
out to be viable~-- an outcome that was not at all clear at the beginning.

Our language is truly domain-specific: for example, it offers no
abstraction mechanism of its own and no general monadic interface for
writing effectful computations. As an upside, the DSL admits useful
non-standard implementations. \S\ref{s:no-do} shows three. In
particular, \S\ref{s:ai} describes an abstract interpretation, to
statically estimate the degree of non-determinism of a DSL term. The
code-generation interpretation~-- the DSL compiler~-- is presented in
\S\ref{s:codegen}. 

\S\ref{s:objections} discusses the presented factoring approach,
answering several commonly heard objections. The main theoretical
ideas have all been known in isolation, often for many decades, as we
review in \S\ref{s:related}.

The source code of all our examples is available at
\url{http://okmij.org/ftp/tagless-final/nondet/}.

\section{Non-determinism through a DSL}
\label{s:list-perm}

This section introduces the DSL that lets us program
all permutations of a given list of integers
in the same starkly elegant way it is done in Curry:
\begin{code}
perm = foldr insert []
\end{code}
This running example, although rather simple (and hence easy to explain
and thoroughly examine), distills large, practical 
projects such as machine-learning \cite{performance-DSL} or the conduct of
clinical trials \cite{tagless-final-bioinformatics}. 
The example is also interesting because
it deals with a rather complex effect~-- non-determinism, which is
rarely supported natively. Yet we are able to use non-determinism just
as easily as in the language Curry, specifically 
designed for non-determinism. We manage with only OCaml
at our disposal, 
which may seem unsuitable since it is call-by-value and
has no monadic sugar. 

\S\ref{s:NDet} defines the DSL, \S\ref{s:Perm} writes our example in
it, and \S\ref{s:no-functor} polishes the DSL embedding by overcoming
the ungainly (and objectionable, to some) functors. The
(standard) implementations of the DSL are discussed in
\S\ref{s:list-impl}~-- and three non-standard ones in 
\S\ref{s:no-do}.

\subsection{DSL Definition}
\label{s:NDet}

We start by designing a language just expressive enough for our
problem of computing a list permutation using non-determinism. We
embed this ``domain-specific'' language (DSL) into OCaml in the
tagless-final style. (Instead of OCaml, we could have used any other
ML or ML-like language~-- or Scala or Haskell or Rust.)  
Recall, in the tagless-final style a DSL is defined
by specifying how to compute the meaning of its expressions
\cite{tagless-final-oxford}. The
meaning is represented by an OCaml value of some 
abstract type (such as the types
|int_t| and |ilist_t| below, the semantic domains of integer and
integer list expressions). The meaning of a complex expression is
computed by combining the meanings of its immediate sub-expressions, that
is, compositionally. A language is thus defined by specifying the
semantic domain types and the meaning computations for its syntactic
forms.  These definitions are typically collected into a signature,
such as:
\begin{code}
module type NDet = sig
  type int_t
  val int: int -> int_t

  type ilist_t
  val nil:  ilist_t
  val cons: int_t -> ilist_t -> ilist_t
  val list: int list -> ilist_t 

  val recur:     (* $\textsf{recur c n lst}$: see text for the explanation *)
    (int_t * ilist_t -> (unit -> ilist_t) -> ilist_t) -> ilist_t -> ilist_t ->
    ilist_t

  val fail:  ilist_t
  val (|||): ilist_t -> ilist_t -> ilist_t
end
\end{code}

Since we will be
talking about integer lists, we need the integer type |int_t| and
at least the integer literals. Whereas |1| is an OCaml integer literal,
the OCaml expression |int 1| represents a DSL integer literal.
We do not define any operations on integers, since they are not
needed for the problem at hand. They can always be added later.  After
all, the ease of extending the language with new types and operations
is the strong suit of the tagless-final
embedding.

We also need integer lists, with the familiar constructors |nil| and
|cons|. The |list| operation turns an OCaml list into a list in our
DSL: |list [1;2;3]| (compare with |int 1| example above).
Although every DSL list can be expressed through |nil| and |cons|, the
special notation for literal DSL lists is convenient.  

We also need a way to recursively analyze/deconstruct lists. For that
purpose, we introduce the recursor |recur|, inspired by the recursor $R$
for natural numbers in G{\"o}del's System T (see Tait
\cite{Tait-intensional} for the modern exposition; Tait calls
$R$ an iteration). Similarly to $R$, the meaning of
our |recur| is specified by the following equalities (or, algebraic identities):
\begin{equation}
\begin{eqnalign}[Tr>{$\;\;\equiv\;\;$}Tl]
|recur $c$ $n$ nil| & $n$\\
|recur $c$ $n$ (cons $h$ $t$)| &
  |$c$ ($h$,$t$) (fun () -> recur $c$ $n$ $t$)|
\end{eqnalign}\label{e:recur}
\end{equation}
As in high-school algebra, an identity states that the two terms
connected by the |===| sign are to be considered `the same'. If an
identity contains variables (such as $c$, $n$, etc. above~-- typeset
in the mathematical font), it should hold for all instantiations,
i.e., replacements of a variable with a term of a suitable type.
One may bet that the thunk |fun () -> ...| visible in \eqref{e:recur}
was not present in 
G{\"o}del's formulation of $R$. Why we have introduced it in our
|recur| will become clear in the next section.

Finally, the |NDet| signature defines the operations for
non-determinism: failure and
the binary choice. Specifically,  \lstinline{l1 ||| l2} denotes a 
non-deterministic choice among
two lists, |l1| and |l2|. To make sure 
the operations \lstinline{l1 ||| l2} and |fail|, however they may end up
being implemented, agree with the intuitions about the non-deterministic
choice and failure, we impose the following identities:
\begin{equation}
\begin{eqnalign}[Tr>{$\;\;\equiv\;\;$}Tl]
|cons $x$ fail| & |fail|
\\
|cons $x$ ($l_1$ CCC $l_2$)| & 
|cons $x$ $l_1$ CCC cons $x$ $l_2$|
\\
|recur $c$ $n$ fail| & |fail|
\\
|recur $c$ $n$ ($l_1$ CCC $l_2$)| &
|recur $c$ $n$ $l_1$ CCC recur $c$ $n$ $l_2$|
\\
|($x$ CCC $y$) CCC $z$| &
|$x$ CCC ($y$ CCC $z$)|
\end{eqnalign}\label{e:choice}
\end{equation}

Any implementation of |NDet| is supposed to verify that the above
identities hold for that implementation. In OCaml, we cannot check the
satisfaction mechanically; we cannot even attach these identities
to the signature except in comments. Wirsing's survey 
\cite{Wirsing-specifications} cites many systems which do verify
the satisfaction of equational specifications.

An attentive reader may get the feeling that something is amiss: the
|NDet| DSL does not look at all like a functional language. There are
no function types (only integers and integer lists) and hence no
operations to construct, or even apply, functions. |NDet| is \emph{not}
a lambda-calculus. How useful can such a trivial language be? On the
other hand, isn't |recur| a higher-order function, from the look of
the type of its first argument? Please hold your
wonder.

%% \begin{exercise}
%% "Our DSL is not the lambda calculus; it is essentially first order"
%% Previously it was said that |foldr| could be defined from what was shown
%% above the point of its definition.  Is there a contradiction here?
%% (We thank the anonymous reviewer for this Koan.)
%% Hint: see Ex.~\ref{e:foldr}
%% \end{exercise}

\begin{exercise}\label{e:foldr}
When one hears about recursively deconstructing a list, what is likely
to spring to mind is |foldr|. Yet for some reason we introduced the relatively
obscure |recur| instead. Can you venture a guess why we did that?
How does |recur| relate to |foldr|?
\end{exercise}
\begin{exercise}
Does it make sense to define separate types for values and expressions
of our DSL? What benefits may come from this separation?
\end{exercise}
\begin{exercise}\label{e:algebra}
The signature |NDet| is not algebraic (why?). 
How would you characterize it?
\end{exercise}
\begin{exercise}\label{e:nondet-laws}
The identities \eqref{e:choice} are by no means the complete
equational specification of non-determinism. What other identities
with |fail| and |CCC| could be added to \eqref{e:choice}?
\end{exercise}

\subsection{List permutation, Non-deterministically}
\label{s:Perm}

However feeble our |NDet| DSL may be, it is enough for the task at hand.
We now use it to write the list permutation as elegantly
as in Curry. For reference, here is the permutation code as it appears
in the Curry standard library:
\begin{code}
insert x [] = [x]
insert x (y:ys) = (x:y:ys) ? (y:insert x ys)
 
perm = foldr insert []
\end{code}

To realize this code in the |NDet| DSL, we first tackle 
the non-deterministic list insertion: |insert x lst| is
to insert the element |x| \emph{somewhere} in |lst|, 
returning the extended list.
That is, it inserts |x| at the front of |lst|, or after the first element of
|lst|, or after the second element of |lst|, etc. The algorithm can be
formulated, and hence implemented, inductively: |insert x lst| either
inserts |x| at the front of |lst| or within |lst|, i.e.,
somewhere in its tail.
Computing the list permutation is now accomplished. The following is the
complete code written in the |NDet| DSL, 
which also includes a simple test.\footnote{The
  accompanying code includes many more (regression) tests.}
\begin{code}
module Perm(S:NDet) = struct
  open S

  (* val foldr: (int_t -> ilist_t -> ilist_t) -> ilist_t -> ilist_t -> ilist_t *)
  let foldr c = recur (fun (h,_) r -> c h (r ()))

  let insert x = 
   recur (fun (h,t) inserted -> cons x (cons h t) ||| cons h (inserted ()))
          (cons x nil)

 let perm = foldr insert nil
 let test1 = perm (list [1;2;3])
end
\end{code}
The DSL primitives such as |recur|, |cons|, |nil| etc. are all defined
in the implementation |S| of the signature |NDet|. The code does not depend
on any particular implementation, which is hence abstracted over as an
argument |S|. Therefore, the DSL code is typically 
represented as an OCaml functor,
parameterized by the DSL implementation (there are nicer-looking
representations, please wait till \S\ref{s:no-functor}). 
Since |NDet| only provides
|recur| but no |foldr|, first we have to implement the latter (with
the expected inferred type shown in the comment). The implementation
is straightforward. The |insert| is straightforward as well, mirroring
the Curry code (keeping in mind that the nondeterministic-choice
operator is spelled |?| in Curry and \lstinline{|||} in our code). The
code keeps the invariant that |inserted ()|, denoting the recursive
invocation of |insert|, is the expression returning the list with
exactly one |x| inserted somewhere. The same invariant is true of the
Curry code.

Although our code looks like the Curry code and is exceedingly simple,
there is something odd about it. We have said that |NDet| has no functions:
no function types, no way to create or apply functions.
What is |insert| then?
Isn't |foldr| a higher-order function? They are functions~--
in the \emph{metalanguage}, rather than in |NDet|. We use the higher-order
facilities of OCaml to construct 
first-order DSL expressions. OCaml truly acts as a
preprocessor for the DSL; |insert| is hence
a `macro'. Our code then is a combination
of a trivial, non-deterministic DSL with a very expressive,
higher-order `macro' system.\footnote{\relax
An old joke comes to mind:
``Much of the power of C comes from having a powerful preprocessor.
The preprocessor is called a programmer.'' \cite{Against-C}.}
Moreover, the DSL evaluation and the `macro-expansion' run like coroutines.
It is not unheard of: after all, coroutines were invented
as a communication mechanism among phases of a Cobol compiler 
\cite{Conway-coroutine}. The coroutining between a
lambda-calculus--based `metalanguage' and the embedded `effectful'
language is the essence of Reynolds' Idealized Algol \cite{reynolds-essence} and
Moggi's computational calculus \cite{moggi-multi-stage}.

\begingroup
\def\I#1{\ensuremath{\overline{#1}}}
\def\L#1{\overline{\mathsf{[#1]}}}
To get a better feeling for the ``macro-expansion'' and also the
confidence in the DSL, it is worth doing a simple exercise:
determine the DSL terms that should be identical to
|perm (list [1;2;3])|. Below we do a part of the exercise, working
out the identities of |insert (int 1) (list [2;3])|.
For the sake of readability, we write DSL terms like |int 1|
as $\I1$ and DSL list literals like |cons (int 2) (cons (int 3) nil)|
as $\L{2;3}$.

\begin{code}
insert $\I1$ $\L{2;3}$
=== (* inlining definitions: ``macro$\hyp$expansion'' *)
recur (fun (h,t) inserted -> cons $\I1$ (cons h t) ||| cons h (inserted ())) $\L{1}$ $\L{2;3}$
=== (* identities $\eqref{e:recur}$ *)
(fun (h,t) inserted -> cons $\I1$ (cons h t) ||| cons h (inserted ())) ($\I2$,$\L{3}$) 
    (fun () -> recur (fun (h,t) inserted -> ...) $\L{1}$ $\L{3}$)
=== (* substitution of values *)
cons $\I1$ (cons $\I2$ $\L{3}$) ||| cons $\I2$ (recur (fun (h,t) inserted -> ...) $\L{1}$ $\L{3}$)
=== (* convention for the literal lists *)
$\L{1;2;3}$ ||| cons $\I2$ (recur (fun (h,t) inserted -> ...) $\L{1}$ $\L{3}$)
=== (* once again identities $\eqref{e:recur}$ *)
$\L{1;2;3}$ ||| cons $\I2$ (cons $\I1$ (cons $\I3$ $\L{}$) |||  cons $\I3$ (recur (fun (h,t) inserted -> ...) $\L{1}$ $\L{}$))
=== (* and again *)
$\L{1;2;3}$ ||| cons $\I2$ ($\L{1;3}$ ||| $\L{3;1}$)
=== (* identities $\eqref{e:choice}$ *)
$\L{1;2;3}$ ||| ($\L{2;1;3}$ ||| $\L{2;3;1}$)
\end{code}
The identities ought to hold in any implementation of |NDet|. Thus, whatever
the implementation, |insert $\I1$ $\L{2;3}$| should amount to the
choice among $\L{1;2;3}$, $\L{2;1;3}$ and $\L{2;3;1}$, in full
agreement with our intuitions.

We have used |===| to mean the least equivalence relation that
contains the identities \eqref{e:recur} and \eqref{e:choice}, and is
closed under substitutions of OCaml values into OCaml lambda-terms. In
other words, |===| includes the ``macro-expansion'' performed as the
ordinary OCaml call-by-value evaluation. The thunk |fun () -> recur $c$ $n$ $t$|
in \eqref{e:recur} was needed precisely for the sake of this
value substitution.
\endgroup
\begin{exercise}Complete the exercise and work out
|perm (list [1;2;3])|.
\end{exercise}

\subsection{Smoother DSL Embedding}
\label{s:no-functor}

One often hears the complaint that writing DSL
expressions as functors is cumbersome.  But there are other
ways, blending the DSL code into the regular OCaml. The result looks
quite like the Lightweight Modular Staging (LMS) in Scala \cite{LMS}~-- the
metaprogramming, DSL-embedding framework which has
been used for `industrial-strength' DSLs.

As a warm-up, let us take one particular DSL implementation, such as |NDetL|
to be described in \S\ref{s:list-impl}. 
Let us write |perm| without any functors this time,
as an ordinary OCaml function:
\begin{code}
let perm : int list -> int list list = fun l ->
  let open NDetL in
  let foldr c = recur (fun (h,_) r -> c h (r ())) in
  let insert x = 
   recur (fun (h,t) r -> cons x (cons h t) ||| cons h (r ())) (cons x nil)
  in foldr insert nil (list l)
\end{code}
This |perm| is truly an ordinary OCaml function, to be applied as
|perm [1;2;3]|.

We now abstract over the DSL implementation. First, we add to |NDet| the
observation operation, so we may \emph{generically} extract the
the list of permutation choices from the result of the |perm|
computation. (One may argue that such a |run| operation should have been
a part of |NDet|. On the other hand, we shall demonstrate non-standard
interpretations of |NDet|, whose results are not
permutation lists but rather static analyses of the generated code.)
\begin{code}
module type NDetO = sig
  include NDet
  val run : ilist_t -> int list list
end
\end{code}
The permutation function will receive the DSL implementation as the
(first-class) module argument:\footnote{The
right-associative infix operator \textsf{@@} of low
precedence is application: \textsf{f @@ x + 1} is the same as
\textsf{f (x + 1)} but avoids the parentheses. The operator is the
analogue of \textsf{\char`$} in Haskell.} %% $
\begin{code}
let perm : (module NDetO) -> int list -> int list list = fun (module S:NDetO) l ->
  let open S in
  let foldr c = recur (fun (h,_) r -> c h (r ())) in
  let insert x = 
   recur (fun (h,t) r -> cons x (cons h t) ||| cons h (r ())) (cons x nil)
  in run @@ foldr insert nil (list l)
\end{code}
Modular implicits \cite{modular-implicits} can even save us the trouble of
passing the |NDet| implementation explicitly. 
DSLs become convenient:
DSL primitives look like the ordinary OCaml operations,
but can be distinguished by their types. Instead of first-class
modules we could have used plain records. Our
approach therefore easily applies to other ML(-like) languages.

\section{Implementing Non-determinism}
\label{s:list-impl}

To run the |Perm| code we need an implementation of the |NDet|
signature. There are many of them, even in this paper (see the
exercises at the end of the section, and \S\ref{s:no-do}). We start
with the `list of successes', the most familiar model of
non-determinism, envisioned already by Rabin and Scott in the
1950s \cite{rabin-scott}. This model is also called `list monad';
our code, however, does not use the monad in its full generality,
as we shall see soon. The realizations of |NDet| in 
\S\ref{s:no-do} cannot be expressed as monads at all, to be explained
there.

In this list implementation, to be called |NDetL|, 
|ilist_t| is the list
of all choices that a list DSL expression may produce:
\begin{code}
type int_t = int
type ilist_t = int list list
\end{code}
We are talking about OCaml lists, which are finite and `eager'. Generally, this
is not the best choice for performance; however, this realization fits
very well our running example, which is to compute the list of all
possible permutation choices. (We shall encounter this interplay of
generality and specialization many more times.) Again, we are
interested in non-deterministic computations on integer
lists only; DSL integers are always deterministic and
therefore, can be represented as plain OCaml |int|. All in all, 
the |NDetL| implementation is as follows:
\begin{code}
module NDetL = struct
  type int_t = int
  let int x = x

  let concatmap: ('a -> 'b list) -> 'a list -> 'b list =
    fun f l -> List.concat @@ List.map f l

  type ilist_t = int list list
  let nil = [[]]
  let cons: int_t -> ilist_t -> ilist_t = 
    fun x -> List.map (fun l -> x::l)
  let list x = [x]

  let rec recur: (int_t * ilist_t -> (unit -> ilist_t) -> ilist_t) -> ilist_t -> ilist_t -> ilist_t =
    fun f z -> concatmap @@ function
      | []   -> z
      | h::t -> f (int h, list t) (fun () -> recur f z (list t))

  let fail: ilist_t = []
  let (|||): ilist_t -> ilist_t -> ilist_t = (@)
end
\end{code}

As expected, literal list expressions such as |list| and |nil| 
are deterministic: have exactly one choice of value. On the other
hand, |fail| has none; \lstinline{(|||)} adds up the choices. 
As was said already,
integer expressions are deterministic by design. Although we have
introduced |concatmap| (the `bind' of the list monad) and could have
likewise introduced `return', we do not export them. In the code they
are used only at specific types (namely, integer lists). It is this
property that will let us later write other implementations of
|NDetL|, which are not at all monadic.

With this |NDetL| implementation of |NDet|, the sample test~-- all
permutations of |[1;2;3]|~-- is run as:
\begin{code}
let module M = Perm(NDetL) in M.test1
### [[1; 2; 3]; [2; 1; 3]; [2; 3; 1]; [1; 3; 2]; [3; 1; 2]; [3; 2; 1]]
\end{code}

\begin{exercise}Check that the identities \eqref{e:recur} and
\eqref{e:choice} hold in this implementation.
\end{exercise}
\begin{exercise}
Consider other implementations of |NDet|, in terms
of delimited continuations, the |delimcc| library, or
operating system threads.
\end{exercise}
\begin{exercise}
Add yet another implementation of |NDet|: e.g., using the
free(r) monad. Besides the depth-first search (underlying the
list implementation), try to implement complete search strategies such
as breadth-first search or iterative deepening.
\end{exercise}
\begin{exercise}
Typically, a tagless-final presentation features the type |'a repr|,
a set of OCaml values that represent DSL expressions of the type |'a|.
We have managed to do without |'a repr|. What have we lost?
\end{exercise}
\begin{exercise}
Generalize the |NDet| signature introducing |'a repr| and
implement this language.
\end{exercise}

\section{Advanced non-determinism: Sorting}
\label{s:sorting}

An immediate application of list permutation is sorting:
sorting, by definition, is obtaining a sorted permutation. This
definition, as is, can be written down in our DSL, giving
us the sorting function |sort|. 
It is called `slow sort'~-- one of the benchmarks
of functional-logic programming. Although not usually fast,
it is correct by definition. The actual
performance depends on the implementation and could be quite good
(that is, not requiring exponential time and space, in the length of
the list).

To express sorting we need two more non-deterministic
primitives. Extending a language defined in the tagless-final
style is easy, by adding new definitions and reusing the old ones:
\begin{code}
module type NDetComm = sig
  include NDet
  val rId  : (int list -> bool) -> ilist_t -> ilist_t 
  val once : ilist_t -> ilist_t
end
\end{code}

The operation |rId| is a form of a logical conditional: it imposes a
guard (a predicate constraint) on a non-deterministic expression. It
is hence akin to |List.filter|. The name
is chosen to match the Curry standard library. The primitive |once|
(called |head| in Curry) expresses the so-called
\emph{don't care non-determinism}:
if an expression has several latent choices, |once| picks one of them.

The sorting is written literally as ``a sorted permutation'':
\begin{code}
module Sort(Nd:NDetComm) = struct
  open Nd
  include Perm(Nd)
  let rec sorted = function
    | []  -> true
    | [_] -> true
    | h1 :: h2 :: t -> h1 >= h2 && sorted (h2::t)

  let sort l = once @@ rId sorted @@ perm l
  let tests = sort (list [3;1;4;1;5;9;2])
\end{code}
\begin{exercise}\label{e:sort-meta}
One may say that the sortedness is expressed
`meta-theoretically'. What makes one say that?
\end{exercise}

Extending a DSL implementation is just as easy as extending the language
definition: we just add the code for the new primitives, which are indeed
primitive:
\begin{code}
module NDetLComm = struct
  include NDetL
  let rId  = List.filter
  let once = function [] -> [] | h::_ -> [h]
end
\end{code}
We can really sort: |let module M = Sort(NDetLComm) in M.tests|
\begin{exercise}\label{e:slow-sort}
The slow sort is particularly slow in the shown list implementation of
|NDet|. Why? How to speed it up?
\end{exercise}
\begin{exercise}
Implement other classical non-deterministic puzzles from
     the Curry example library
\url{http://www.informatik.uni-kiel.de/~mh/curry/examples/}
\end{exercise}

\section{When Monads will not \textsf{do}}
\label{s:no-do}

Although the |NDet| DSL is meant for non-deterministic computations,
it is not as generic and expressive as it could be. For example, the
|NDet| signature does not define the general monadic `bind' and
`return' operations (they were not needed for the task at hand).
Implementations of |NDet|, such as |NDetL| in
\S\ref{s:list-impl}, may support these operations and even use
them internally~-- yet not offer them to the DSL programmer. The
lack of generality has an upside: the |NDet| DSL admits
implementations that do not support `bind' and `return' at all. 
This section presents three non-monadic interpretations of
|NDet|, and explains why they are interesting and why they fall
outside the conventional monadic framework.

\subsection{More efficient representation}

The |NDet| signature in \S\ref{s:NDet} flaunts the extreme
specialization: the DSL has only integers and integer lists as data
types. The conspicuous lack of general lists admits however
an efficient representation. Rather than the familiar linked
list of cons cells (with each cell holding one list element), we may group
elements in tightly packed chunks, e.g., like in Bagwell's VList
\cite{Bagwell-VLIST}.
A chunk can be represented as an array, or even 
OCaml's |Bigarray|.\footnote{\relax
\url{https://caml.inria.fr/pub/docs/manual-ocaml/libref/Bigarray.html}} The
latter is particularly efficient, e.g., in avoiding GC
marking. However, Bigarrays are not polymorphic: they are restricted
to integers and floating-point numbers. It so happens our integer
lists fit the restriction. The accompanying code shows an
implementation of |NDet| where integer lists are (very naively, at
present) represented with |int Bigarray| chunks. It is now an
advantage that the |NDet| signature fails to define |return| and |bind|,
because we would not have been able to support them:
the present implementation deals with non-deterministic
computations on integer lists only.

Again we see the general/specific trade-off: restricting the
expressivity (the set of data types to operate upon) may gain a more
efficient data representation.

\subsection{Abstract Interpretation}
\label{s:ai}

The |NDet| DSL signature admits truly non-standard
interpretations. This section describes one such example:
instead of actually performing a non-deterministic
computation, we estimate the number of its non-deterministic choices
and the possibility of failure. This is an example of the static analysis
known as abstract interpretation
\cite{CousotCousot77,JonesNielson95}. 
Our example is realistic: the Kiel Curry compiler, for one, performs a
similar determinism analysis in order to produce efficient code
\cite{KiCS2}.

Recall, the tagless-final DSL is defined by specifying how to
compositionally compute the meaning of its expressions. The `standard'
interpreter such as |NDetL| in \S\ref{s:list-impl} takes the meaning
of a non-deterministic expression to be the set 
(to be more precise, the OCaml list)
of possible values. The
non-standard interpreter to be developed in this section uses a coarse,
`abstract', semantic domain, merely approximating that set.
Namely, our abstraction domain here 
is the expression's degree of non-determinism:
\begin{code}
type ndet_deg = {can_fail: bool; choices: Iint.t}
\end{code}
It records the possibility of failure and the upper bound on the number
of possible values: one, two, three, etc., or many. (See
Fig.\ref{f:Iint} for one implementation of integers with `many'.)
An expression of
degree |d1| is at least as non-deterministic as an expression of
degree |d2| (written as |d2 $\le$ d1|) iff
\begin{code}
d2.choices <= d1.choices $\land$ d2.can_fail <= d1.can_fail
\end{code}
In fact, our domain is not just a partial order but a join
semi-lattice (common in abstract interpretation), so that we can
compute the least-upper bound on the non-determinism degree for any
set of expressions (for binary joins, see |join| in
Fig.\ref{f:NDetAbsND}).
\begin{figure}
\begin{code}
module Iint = struct
  type t = Int of int | Inf          (* integers with infinity *)

  let one  : t = Int 1
  let zero : t = Int 0
  let inf  : t = Inf

  let ( + ) : t -> t -> t = fun x y ->
    match (x,y) with
    | (Int x, Int y) -> Int (x+y)
    | _ -> Inf

  let ( * ) : t -> t -> t = fun x y ->
    match (x,y) with
    | (Int x, Int y) -> Int (x*y)
    | _ -> Inf

  let ( <= ) : t -> t -> bool = fun x y ->  (* partial order *)
    match (x,y) with
    | (_,Inf) -> true
    | (Inf,_) -> false
    | (Int x, Int y) -> x <= y

  let max : t -> t -> t = fun x y ->    (* join of the semilattice *)
    match (x,y) with
    | (Int x, Int y) -> Int (if x > y then x else y)
    | _              -> Inf
end
\end{code}
\caption{A semiring/join semilattice with `many' (\textsf{inf})}
\label{f:Iint}
\end{figure}

Since the degree of non-determinism is estimated statically, before
evaluating an expression, it is an approximation. It is
an over-approximation: an expression with the estimated degree
|{can_fail=true;| |choices=Int 5}| may in reality finish without
failure, with only two possible values. Our interpreter
however guarantees that the over-approximation is sound: the 
expression in question may have fewer than 5 possible values, but not
more than 5. An expression with the (smallest) degree:
\begin{code}
let det = {can_fail=false; choices=Iint.one}
\end{code}
is therefore certainly deterministic. The largest degree (the maximal
element of the domain) is
\begin{code}
let top = {can_fail=true; choices=Iint.inf}
\end{code}
It is the least informative estimate of the actual degree of non-determinism.

The abstraction domain has more structure than a mere join
semi-lattice. If |e1| is the non-deterministic list (expression) with
at most 2 choices and |e2| is the expression with at most 3 choices,
we would like to estimate that \lstinline{e1 ||| e2} has at most 5
choices. On the other hand, concatenating the lists |e1| and |e2|
should have at most 6 choices of the result.
We thus need additive and multiplicative
operations. Whereas the former is used only in
interpreting \lstinline{(|||)}, the multiplicative operation is more
common. It is called |merge| in the code in Fig.\ref{f:NDetAbsND};
incidentally, |det| is its unit:
|merge det d === d| for any degree |d|.

Figure \ref{f:NDetAbsND} shows the complete code of the abstract
interpreter, most of which has
already been explained. 
As any other DSL interpreter, the abstract interpreter also 
implements the signature |NDet|. 
The meaning of an |ilist_t| expression is its degree of
non-determinism; |int_t| expressions, which are always deterministic,
are represented by an abstract integer |AInt|. The interpretation 
of the recursor deserves a few words. Since our abstraction domain
keeps track only of the degree of nondeterminism for a list expression
but not of the length of the list, the best we can do to approximate
|recur c n l| is to find the upper bound for |recur c n| when applied
to lists of every possible length:
\begin{code}
$\bigsqcup_{i=0}$ recur c n [AInt]$^i$
\end{code}
where [AInt]$^i$ is the list of length |i| made of abstract integers.
The recursive equations defining |recur| make it easy to compute
|recur c n [AInt]$^{i+1}$| if
|recur c n [AInt]$^{i}$| is known. All that remains is to keep 
joining until the
result `stabilizes'. Since we are computing an approximation of the
degree of nondeterminism, we would be satisfied with an upper bound,
not necessarily the least one. Therefore, we can stop the joining
iteration after some number of steps, returning |top| if the
convergence has not been achieved by then.
One may organize the fixpoint computation
differently: Abstract Interpretation is a vast area. The presented
tagless-final framework helps us experiment with such static analyses.

\begin{figure}
\begin{code}
module NDetAbsND = struct
  type ndet_deg = {can_fail: bool; choices: Iint.t}

  let det = {can_fail=false; choices=Iint.one} (* deterministic computations *)
  let top = {can_fail=true; choices=Iint.inf}

  let merge : ndet_deg -> ndet_deg -> ndet_deg = fun d1 d2 ->
    {can_fail = d1.can_fail || d2.can_fail;
     choices = Iint.(d1.choices * d2.choices)}

  let join : ndet_deg -> ndet_deg -> ndet_deg = fun d1 d2 ->
    {can_fail = d1.can_fail || d2.can_fail;
     choices = Iint.(max d1.choices d2.choices)}

  type int_t = AInt                     (* An abstract integer *)
  let int: int -> int_t = fun _ -> AInt

  type ilist_t = AList of ndet_deg
  let nil:  ilist_t = AList det
  let cons: int_t -> ilist_t -> ilist_t = fun _ x -> x
  let list: int list -> ilist_t   = fun _ -> AList det

  let merge_lst : ilist_t -> ilist_t -> ilist_t = fun (AList d1) (AList d2) ->
    AList (merge d1 d2)
  let join_lst : ilist_t -> ilist_t -> ilist_t = fun (AList d1) (AList d2) ->
    AList (join d1 d2)

  let recur:
    (int_t * ilist_t -> (unit -> ilist_t) -> ilist_t) -> ilist_t -> ilist_t ->
    ilist_t = fun f z l -> 
        let rec loop acc res_i i =
          let res_i' = f (AInt, AList det) (fun () -> res_i) in
          let acc' = join_lst acc res_i' in
          if acc = acc' then acc 
          else if i > 5 then AList top 
          else loop acc' res_i' (i+1)
        in merge_lst l (loop z z 0)

  let fail: ilist_t = AList {can_fail=true; choices=Iint.one}
  let (|||): ilist_t -> ilist_t -> ilist_t = fun (AList d1) (AList d2) ->
    AList {can_fail = d1.can_fail && d2.can_fail;
             choices = Iint.(d1.choices + d2.choices)}
end
\end{code}
\caption{Abstract interpreter to estimate the degree of
  non-determinism}
\label{f:NDetAbsND}
\end{figure}

Let's take a few examples of the non-determinism analyses:
\begin{code}
let open NDetAbsND in  cons (int 20) (nil ||| cons (int 10) nil) ||| fail
###- : NDetAbsND.ilist_t =
NDetAbsND.AList {NDetAbsND.can_fail = false; choices = NDetAbsND.Int 3}
\end{code}
The result tells that the given non-deterministic list computation, if
evaluated, will have at most three possible values, and it will
not fail.  The result of evaluating
\begin{code}
let open NDetAbsND in foldr cons nil (list [1;2;3])
\end{code}
shows that the |foldr| expression is deterministic. The first argument
of |foldr| (or |recur|) may ignore its arguments:
\begin{code}
let open NDetAbsND in recur (fun _ _ -> fail) (list [1] ||| list [2]) nil
### - : NDetAbsND.ilist_t =
NDetAbsND.AList {NDetAbsND.can_fail = true; choices = Iint.Int 2}
\end{code}

It is easy to see from the abstract interpreter code that we never
underestimate the degree of non-determinism. Thus the analysis is
sound. As an example:
\begin{code}
let open NDetAbsND in
foldr (fun x l -> l ||| cons x l) nil (list [1;2;3])
### - : NDetAbsND.ilist_t =
NDetAbsND.AList {NDetAbsND.can_fail = true; choices = NDetAbsND.Inf}
\end{code}
We can even analyze the permutation code (see \S\ref{s:Perm})
\begin{code}
let module M = Perm(NDetAbsND) in NDetAbsND.observe M.test1
\end{code}
(with the same outcome).

Finally, we should stress that we would not have been able to
abstractly interpret the DSL code, had the |NDet| signature required
the monadic operation |bind|, whose general signature, recall, is
\begin{code}
 val bind   : 'a m -> ('a -> 'b m) -> 'b m
\end{code}
for some parameterized type |m|. The second argument to |bind|, 
the continuation, is to receive the value |'a| produced by the 
computation of the first
|bind| argument. If |'a m| is realized as  
|ndet_deg|, it can never produce any concrete |'a| value.
Therefore, when abstractly interpreting the |bind| expression,
we cannot ever invoke, and hence analyze, its continuation. 
That monadic programs cannot be statically analyzed by choosing a
suitable abstract monad interpretation was the main motivation
for applicative functors \cite{Applicative}
and arrows \cite{Arrows}. 
We refer to that literature for more discussion.

\begin{exercise}
The |recur| code in Fig.\ref{f:NDetAbsND} stops the joining after 5
iterations. Explain why stopping after the first iteration would have
sufficed.
\end{exercise}
\begin{exercise}
Make the analysis more precise by also tracking the size of the 
integer list, if known statically.
\end{exercise}
%% \begin{exercise}
%% Extend the |NDet| signature with operations to build
%% an `unknown' list or an integer
%% \end{exercise}

\subsection{Code Generation}
\label{s:codegen}

This section describes yet another interpreter for the |NDet| DSL,
which is non-standard in a different way. Rather than evaluating a DSL
expression, it generates code for it. The code can be saved into a
file, and then compiled and linked as any other OCaml code. The
interpreter in this section is thus a DSL compiler, turning DSL
expressions into ordinary OCaml code and libraries.

For code generation we rely on MetaOCaml \cite{MetaOCaml}, which is a
superset of OCaml that adds the type |'a code| denoting so-called code
values: (fragments of) the generated code. MetaOCaml provides
two primitives to build such code values. Brackets quote an OCaml
expression
\begin{code}
let c = .<5 + 7>.
### val c : int code = .<5 + 7>. 
\end{code}
turning it, \emph{without evaluating}, into a fragment of the 
generated code. The \emph{escape}, or splice, is a form of
antiquotation, in Lisp terminology. It lets us build code templates
with holes in them, to be later filled with other fragments, for example:
\begin{code}
let template x y = .<if .~x>1 then .~y else .~y*2>. (* the template with two holes, x and y *)
### val template : int code -> int code -> int code = <fun>

template .<read_int ()>. c   (* c was defined in the previous example *)
### - : int code = .<if (Stdlib.read_int ()) > 1 then 5 + 7 else (5 + 7) * 2>. 
\end{code}
Clearly, the code value (the generated code it contains) 
can be printed. It can also be saved into a file. The MetaOCaml home
page \cite{MetaOCaml} has more examples and explanations, with pointers to
various tutorials.

The DSL compiler code in Fig. \ref{f:codegen} interprets DSL
expressions as code values: the fragments of code, which, when compiled and
executed as part of the complete program will compute the expression
values. For example, for integer DSL expressions we have:
\begin{code}
type int_t = int code
let int x = .<x>.
\end{code}
We could have represented list DSL expressions likewise, as the code
to compute the list of all choices:
\begin{code}
type ilist_t = int list list code
\end{code}
The experience with abstract interpretation has taught us to analyze,
to find out what we can say about the program before running it.
We therefore incorporate some analysis (typically called 
`binding-time analysis' \cite{jones-partial})
into the DSL compiler, which
calls for the more elaborate 
semantic domain for non-deterministic list expressions:
\begin{code}
type ilist_t = 
    | K of int list code list
    | U of int list list code
\end{code}
It distinguishes the case of statically knowing the number of
non-deterministic choices~-- in particular, knowing that a list
expression is in fact deterministic. The literal list expression such
as |list [1;2;3]| is clearly deterministic. We note that fact (by
representing it with the |K| variant) and use later on in code
generation (see Fig.\ref{f:codegen}). 
The |U| variant of |ilist_t| corresponds to the
statically unknown degree of non-determinism. It contains the code
computing the choices at run-time. In contrast, in the |K| variant
the choices are known statically, although the content
of each choice is generally not and is to be computed at
run-time. |U| and |K| hence act as annotations on the generated
code~-- so-called \emph{binding-time} annotations. The annotations can
be erased: one may always forget the static
knowledge and return the opaque |int list list code| value. That is
the purpose of the function |dyn| in Fig.\ref{f:codegen}. Among other
uses, it extracts the result of compiling the DSL expression.

Most of the DSL compiler is derived from the list monad implementation
|NDetL| in \S\ref{s:list-impl} by placing brackets and escapes at
appropriate places. (The module type |lift| and its implementations in
|Lifts| are provided by MetaOCaml to `lift' OCaml values to
the code that, when later run, will produce that value. Lifting is
possible only for selected OCaml types.)

The recursor again needs a bit of explaining. Recall that in an
expression |recur c n l|, |l| is a non-deterministic list computation.
Therefore, the |recur| compiler in Fig.\ref{f:codegen} starts by
checking what is already known about |l|: if it is definitely
the failed
computation (in which case the whole |recur| expression is also a
failure), or if it is deterministic. In the latter case, we get
|recur1| to handle its only choice. In the general case, we build the
code to process (again, using |recur1|) all the choices that |l| could
produce, when evaluated.

\begin{figure}
\begin{code}
module NDetLCode = struct
  type int_t = int code
  let int x = .<x>.

  (* Utilities *)
  let scons : 'a code -> 'a list code -> 'a list code = fun x l ->
    .<.~x :: .~l>.
  let concatmap: ('a -> 'b list) code -> 'a list code -> 'b list code =
    fun f l -> .<List.concat (List.map .~f .~l)>.
  
  type ilist_t = 
    | K of int list code list
    | U of int list list code

  let dyn : ilist_t -> int list list code = function
    | K ls -> List.fold_right scons ls .<[]>.
    | U ll -> ll

  let nil = K [.<[]>.]
  let cons: int_t -> ilist_t -> ilist_t = 
    fun x -> let x = genlet x in function
      | K ll -> K (List.map (scons x) ll)
      | U ll -> U .<List.map (fun l -> .~(scons x .<l>.)) .~ll>.
  let list x =                          (* Lifts is part of MetaOCaml *)
    let open Lifts in let module M = Lift_list(Lift_int) in
    K [M.lift x]

  let recur1: (int_t * ilist_t -> (unit -> ilist_t) -> ilist_t) -> 
            ilist_t -> int list code -> ilist_t = fun f z l -> 
      U .<let rec go = function
          | []   -> .~(dyn z)
          | h::t -> .~(dyn @@ f (.<h>.,K [.<t>.]) (fun () -> U .<go t>.))
          in go .~l>.

  let recur: (int_t * ilist_t -> (unit -> ilist_t) -> ilist_t) -> ilist_t -> ilist_t -> ilist_t =
    fun f z -> function
      | K []  -> K []
      | K [l] -> recur1 f z l
      | ls     ->  U (concatmap .<fun l -> .~(dyn @@ recur1 f z .<l>.)>. (dyn ls))

  let fail: ilist_t = K []
  let (|||): ilist_t -> ilist_t -> ilist_t = fun l1 l2 -> match (l1,l2) with
    | (K l1, K l2) -> K (l1 @ l2)
    | (K ls, U ll)
    | (U ll, K ls) -> U (List.fold_right scons ls ll)
    | (U l1, U l2) -> U .<.~l1 @  .~l2>.

  let obs : ilist_t -> int list list code = dyn (* Finally, the observation function *)
end
\end{code}
\caption{The staged DSL interpreter: the DSL compiler}
\label{f:codegen}
\end{figure}

For the sample |Perm.test1| in \S\ref{s:Perm} we generate the 
following code
\begin{code}
val pcode : int list list code = .<
  let lv_10 = [1; 2; 3] in
  let rec go_11 = function
    | [] -> [[]]
    | h_12::t_13 ->
        Stdlib.List.concat @@
          (Stdlib.List.map
             (fun l_14 ->
                let rec go_15 = function
                  | [] -> [[h_12]]
                  | h_16::t_17 -> (h_12 :: h_16 :: t_17) ::
                      (Stdlib.List.map (fun l_18 -> h_16 :: l_18)
                         (go_15 t_17)) in
                go_15 l_14) (go_11 t_13)) in
  go_11 lv_10>. 
\end{code}
When compiled and run, it produces the list of all permutations of the
given sample list |[1;2;3]|. The code is surprisingly clear; one could
have written something like it by hand.

\begin{exercise}Check that the identities \eqref{e:recur} and
\eqref{e:choice} hold in this implementation as well.
\end{exercise}
\begin{exercise}
Think about the ways to improve the binding-time analysis. For
example, how to represent the choices that are only partially
statically known? The list to process may also be (fully or partially)
known statically. When unrolling recursive calls, beware of code
explosion.
\end{exercise}
\begin{exercise}\label{e:sort-code}
How would you extend the DSL compiler to generate slowsort code,
explained in \S\ref{s:sorting}?
\end{exercise}

This DSL compiler would not have been possible had the |NDet| DSL 
required the monadic interface. Indeed, if it were, it would have had to support
the following operations:
\begin{code}
 val return : 'a -> 'a code
 val bind   : 'a code -> ('a -> 'b code) -> 'b code
\end{code}
Both of them are deeply problematic. First, not every OCaml value is
convertible to the code that can be saved into a file and, when run,
reproduces the value. Think of 
closures, reference cells and I/O channels: which code to write into a
file to represent the currently open I/O channel in its current state?
We are \emph{not} saying that the types |(int->int) code|, |int ref code|,
|in_channel code|, etc. are unpopulated. They clearly are, for example
|.<ref 1>. : int ref code|. What we cannot do is to take an |int ref|
value (a location in the current program heap) and convert it to
code to save into a file, which, when compiled and run will yield the
same location. After all, by the time the generated code is run the
current program along with its heap may be long gone. The purpose of
the |lift| module briefly mentioned earlier is to delineate the types
of those values that can be converted to the corresponding code (i.e.,
liftable). The operation |bind| is likewise problematic for code
values. Its second argument is a function that takes the value
meant to be produced by the code supplied as the first argument to
|bind|. The generated code, generally, cannot be run until the
generation process is finished: for example, because the code may
contain free variables, to be bound later in the process. Therefore,
|bind| cannot in general apply its second argument. To put it another way, code
generation cannot, generally, be influenced by the result of the
already built code. All in all, |bind| and |return| with the above
interface and satisfying the familiar monad laws are inexpressible.

\section{Objections and Discussion}
\label{s:objections}

Having presented the experiment of writing effectful programs as a
combination of a simple DSL embedded into a powerful metalanguage,
we now discuss the results. This section concentrates on the comparison
to monads and answering the commonly heard objections. \S\ref{s:related}
discusses the history and the origins of the underlying
theoretical ideas.

\subsection{Do we still clearly separate effectful computations?}

One of the deservingly appreciated benefits of monadic programs is the
clear separation of effectful computations in types and syntax. Our
DSL is meant to be simple and first-order, and hence does not support
the monadic interface.  Yet, |NDet| exhibits an equally
clear separation of effectful computations. Anything of the type
|ilist_t| is potentially non-deterministic; everything else is
deterministic.  Thus from the type of 
|insert : int_t -> ilist_t -> ilist_t| we immediately tell that 
|insert| deterministically transforms non-deterministic computations.

\subsection{It is not generic}

Another benefit of monads is the uniformity of representing many
(although not all) sorts of effectful computations and especially
effectful abstractions: higher-order effectful functions. Our
factoring approach is not generic. An anonymous reviewer of an early
version of this paper (the extended abstract submitted to the ML Family
Workshop 2017) well described the situation and voiced the objection
as follows:
\begin{quotation}
This style of programming with non-determinism seems both obvious and
awkward. Everything has to be explicitly lifted and insofar as the
approach is different from the traditional monadic way of structuring
this kind of code, it seems less generic and less uniform (there's no
guiding structure to say how higher-order functions should be lifted,
for example).

If so, this caveat should be pointed out, as it stands in contrast to
the situation with the monadic approach, where a single definition of
the monad suffices for all types.
\end{quotation}

One may quibble with the negative tone of ``everything has to be
explicitly lifted'' assessment: after all, in the monadic approach one
also has to lift all literals and the results of any pure
computation. Monadic |return| is ubiquitous. As far as the facts of
the matter are concerned, the reviewer's description is accurate. The
factoring approach is not generic, is not generally applicable, and is
not uniform.

Our experiment has demonstrated, however, that we did not need
higher-order domain-specific functions to successfully solve the
problems at hand. The experience with LMS \cite{LMS} likewise shows
that for many practical problems (including machine learning and data
base queries), a first-order DSL is sufficient. We are not the first
to observe that many typical functional programs can be written
without higher-order functions (we discuss this point in more detail
in \S\ref{s:related}).  The first-order nature of the DSL greatly
simplifies its implementation and reasoning. As to generality, even in
Haskell community one begins to hear the advice ``don't generalize
until you use it twice'' and ``strive for meaningful rather than
generic interfaces''.

The lack of uniformity and of genericity has an upside: 
efficient and non-standard interpretations, as we described
in \S\ref{s:no-do}. In particular, code generation also lacks a
uniform way to lift a value to the corresponding code, see
\S\ref{s:codegen}. That does not pose a problem for our approach but
does for the monadic one.

\subsection{But monads are `pure'!}

Finally, we cannot pass on the commonly heard slogan that the monadic
code is `pure' (and, by implication, is `better'). Purity appeared
in the Christoph H{\"o}ger's message, quoted in \S\ref{s:motivation}, 
that prompted this paper. Purity is often used as a political slogan
and rallying cry.\footnote{\relax
It is worth pointing out that Hughes \cite[\S1]{hughes-why}
noted that the standard advantages of functional programming~--
referential transparency, absence of side effects, no explicit control
flow~-- is something that outsiders do not take too seriously. 
``Even a functional programmer should be dissatisfied with these so-called
advantages,'' he wrote.}
If we do look at the purity of monadic code
rationally, we see nothing but confusion.

Indeed, let's look again at Christoph H{\"o}ger's example,
extended with an extra line for illustration:
\begin{code}
do_;
  i <-- let f x = incr x in f 3 ;
  p <-- put (5 + i) ;
  j <-- let f x = incr x in f 3 ;
  return p
\end{code}
It may make sense to abstract the pattern:
\begin{code}
let putp m = do_;
  i <-- m ;
  p <-- put (5 + i) ;
  j <-- m ;
  return p
\end{code}
The original code is recovered by the instantiation
|putp (let f x = incr x in f 3)|. Suppose |putp| is used as a part of a
bigger computation:
\begin{code}
let big m = do_;
  i <-- m;
  j <-- putp m;
  return (i+1,j)
\end{code}
Since |big| has to evaluate its argument |m| anyway, one may be very
tempted to `optimize' |big| as
\begin{code}
let bigO m = do_;
  i <-- m;
  j <-- putp (return i);
  return (i+1,j)
\end{code}
That is, we share the result of the computation rather than the
computation |m| itself~-- and, inadvertently, change the behavior of
our program. In this simple example, the problem is rather
apparent. On one hand, one should not be too surprised: higher-order
facility~-- just like the C preprocessor~-- gave us the ability to
abstract computations rather than values. Functions such as |big|,
like C macros, can be rather subtle: their seemingly straightforward
refactoring often leads to subtle bugs. To be sure, 
this is not the problem created by monads~-- yet monads do little to
ameliorate it. When we write effectful code~-- monads or no monads~--
we have to constantly keep in mind the context of expressions we pass around.

The fact that monadic code `desugars' (is implementable in terms of)
side-effect-free code is irrelevant. When we use monadic notation, we
program within that notation~-- without considering what this notation
desugars into. Thinking of the desugared code breaks the monadic
abstraction. A side-effect-free, applicative code is normally compiled
to (that is, desugars into) C or machine code. If the desugaring
argument has any force, it may be applied just as well to the
applicative code, leading to the conclusion that it all boils down to
the machine code and hence all programming is imperative.

Like Hughes \cite[\S1]{hughes-why}, I object to the purity argument
also methodologically. A particular programming style should be judged
on its merits rather than on appeal to emotion. The merits (ease of
writing, ease of implementing, code reuse among several
implementations, extensibility) ideally should be evaluated by
observation and experiment. Unfortunately, (properly done) empirical
studies of programming styles are few and far between. From the
personal experience, I have noticed that the mistakes I make when
writing monadic code are exactly the mistakes I made when programming in 
C.\footnote{In fact, the \textsf{bigO} example is a very simple 
version of an actual problem in one of my programs. 
That mistake has lead to the redesign of
the interface of enumerators, making it less elegant but also less
error-prone.} Actually, monadic mistakes tend to be worse, because monadic
notation (compared to that of a typical imperative language) is
ungainly and obscuring\aside{(e.g., state monad code often has to
  pack/unpack the global state, which tends to introduce subtle
  bugs)}.

\section{History and connections}
\label{s:related}

We owe the main idea of the factoring approach~-- representing a
program as a simple DSL with a powerful macro system~-- to Milner and
his Meta Language \cite{ML-original}. We use the metalanguage,
however, to build executable, effectful expressions rather than
formulas and theorems. One may trace the origin of the approach back to Church's
design of the typed lambda-calculus \cite{church-formulation}, meant to be the
metalanguage providing for abstraction, definition and naming~-- into
which one may embed a logic DSL with constants such as equality,
$\lor$, $\forall_\alpha$, etc. This idea was further developed in
the Logical Framework LF \cite{LF}. ML was explicit, however, in letting
programmers define their own interpretations of constants~-- 
what we have demonstrated with several 
different implementations of the |NDet| signature.

That typical higher-order functional programs can be written
in a first-order language enhanced with parameterized modules (that
is, endowed with a good `macro' facility) was clearly enunciated by
Goguen \cite{Goguen-noHO}.
\begin{quotation}
``I do not consider higher order functions \emph{harmful}, 
\emph{useless}, or \emph{unbeautiful}; but I do claim significant
advantages for avoiding higher order functions whenever possible,
and I claim they can be avoided quite systematically in functional
programming, by using parameterized programming instead.''
\cite[Sec. 1]{Goguen-noHO}
\end{quotation}
Avoiding higher-order functions, Goguen pointed out, brings
simplicity and efficiency to interpreters and compilers, and,
mainly, the ease of reasoning: correctness proofs can be done entirely
in first-order logic. The tagless-final style we expound in the
present paper may be considered an instance of Goguen's parameterized
programming. We do not limit the signatures to (conventionally)
algebraic (see Ex.\ref{e:algebra}).

\begin{comment}
The study
of call-by-push-value shows that we can interpret computations using
algebras of an algebraic (or Lawvere) theory. It appears that the paper is
implementing this idea (i.e. algebraic models of computations) inside
OCaml, and the OCaml module system is used to structure the concept of
algebras inside the language. It is natural, but at the same time I
was not very surprised by this approach. I therefore do not recommend
accepting this paper very much.
\textit{Anonymous Reviewer}

Actually, even Barendregt in his well-known book starts with algebraic
semantics.

\oleg{Reynolds, Felleisen and White}
\end{comment}

Moggi and Fagorzi \cite{moggi-multi-stage} described monads as a tool
for structuring~-- staging~-- of effectful computations.  There are
other ways to introduce sublanguages, such as the tagless-final style
shown off in the present paper.

Robert Atkey has pointed out Reynolds' argument in
\cite{reynolds-essence} that Algol is the orthogonal combination of
lambda-calculus and imperative programming. (Later, Abramsky and
McCusker \cite{Abramsky-linearity} described the `imperative
programming' part as an interaction with a process that implements the
behavior of a storage cell.) Lambda-calculus can thus be thought of as
a metalanguage, with the imperative part modeled as the following DSL
(of process-interaction combinators):
\begin{code}
module type STATE = sig
  type comm
  type exp
  type var
  
  val skip : comm
  val seq  : comm -> comm -> comm
  
  val const  : int -> exp
  val add    : exp -> exp -> exp
  val (:=)   : var -> exp -> comm
  val read   : var -> exp
  val while_ : exp -> comm -> comm
  val new_   : (var -> comm) -> comm
end
\end{code}
(I am very grateful to Robert Atkey for this signature and example.)
On this formulation, Algol also has the `programmable semicolon': the
|seq| operation~-- as well as the programmable loop.

Harper \cite[Sec.20. Modalities and Monads in
  Algol]{Harper-commentary} argues that the distinction between
`pure' (context-independent) and effectful
computations is \emph{modal} but not \emph{monadic}: specifically, in a
so-called lax modality \cite{PD-judg,lax-logic}.

\section{Conclusions}

We have described a direct alternative to the monadic
encoding of effects: a bare-bone domain-specific language with
effectful operations.  The DSL is blended into a
metalanguage such as OCaml; therefore, it can be kept tiny, with no
abstraction facilities of its own, or even functions.  The
metalanguage, serving as an inordinarily expressive macro system,
compensates. We have also argued for the principle of avoiding
premature generalizations and abstractions.  

We have reported only one experiment, which~-- combined with the
related LMS experience~-- suggests that the DSL-metalanguage factoring
approach to effectful programming is viable. More experiments are
needed to better grasp its usefulness. Specifically we would like to
try examples in the scope of Async or Lwt libraries. A bigger
exercise would be to re-implement the programming language
Icon~-- another language with built-in non-determinism.

One may wonder how the history of (meta) programming might have turned out
if the ML evolution had taken a different turn: kept using ML as the
Meta Language as it was initially designed for, but building
objects other than formulas and theorems~-- in particular, programs.

\subsubsection*{Acknowledgments}
Extensive comments and suggestions by anonymous reviewers
are greatly appreciated.
I am particularly grateful to
Robert Atkey for very many helpful suggestions, and for
explanations of Idealized Algol.
I thank Robert Harper for pointing out the lax modality and
its discussion.
This work was partially supported by JSPS KAKENHI Grant Number
17K00091.

\bibliographystyle{eptcs}
\bibliography{../../ExtEff/exteff,nondet,mybib}

\appendix
\section{Hints to selected exercises}

\paragraph{List recursors (Ex.~\ref{e:foldr})}
Please try to write the function |tail| (obtaining the tail of a list)
in terms of |foldr| and |recur|.

For more discussion of G{\"o}del recursor and its connection with
the fold on natural numbers (i.e., Church numerals) see
\url{http://okmij.org/ftp/Computation/lambda-calc.html#p-numerals}.

\paragraph{Algebraic specifications (Ex.~\ref{e:algebra})}

To answer the question if |NDet| is algebraic we need the definition
of algebraic specification.
For ease of reference, we quote the definition from
Wirsing's reference article \cite[\S2.1]{Wirsing-specifications}
(see also Burris and Sankappanavar's detailed course 
\cite{UniversalAlgebraCourse}).
Formally, a (multi-sorted algebraic) signature $\Sigma$ is a pair
$\langle S,F\rangle$ where $S$ is a set (of sorts) and $F$ is a set (of function
symbols) such that $F$ is equipped with the mapping 
${type}: F\to S^n \times S$ for some $n\ge 0$. The mapping, for a
particular $f$ of $F$ is often denoted as
$f: s_1,\ldots,s_n\to s$ where $\{s,s_1,\ldots,s_n\} \subset S$.
A $\Sigma$-Algebra consists of an $S$-sorted family
of non-empty (carrier) sets $\{A_s\}_{s\in S}$ and a total
function $f^A: A_{s_1},\ldots,A_{s_n}\to A_s$ for each
$f: s_1,\ldots,s_n\to s \in F$.
For example, in the following OCaml code
\begin{code}
module type NAT = sig
  type nat
  val zero : nat
  val succ : nat -> nat
  val plus : nat -> nat -> nat
end
module Nat : NAT = struct
  type nat = int
  let zero = 0
  let succ x = x + 1
  let plus x y = x + y
end
\end{code}
|NAT| is the signature, whose set of sorts is the singleton
$\{\mathrm{nat}\}$ and function symbols are $\{\mathrm{zero},
\mathrm{succ},\mathrm{plus}\}$. In Wirsing's notation, one would write
the type of |plus| as 
$\mathrm{plus}: \mathrm{nat},\mathrm{nat}\to \mathrm{nat}$. 
|Nat| is a |NAT|-algebra, whose carrier is the set of OCaml
integers.
See also Staton's extension of algebra
formalism \cite{Staton-algebras}, permitting `richer' signatures.

\paragraph{Laws of non-determinism (Ex.~\ref{e:nondet-laws})}
Which equational laws should hold for non-deterministic computations
is a rather complicated and controversial question, with no single
answer. The web page
\url{http://okmij.org/ftp/Computation/monads.html#monadplus} discusses
some of the complexities.

\paragraph{Sortedness, meta-theoretically (Ex.~\ref{e:sort-meta})}
The exercise is an invitation to contemplate once again how the
overall (sorting) computation is spread across the DSL and the
metalanguage. The type of |rId| is particularly worth examining
closely, asking oneself what do |int list| and |ilist_t| represent and
what is the difference between them. See also Ex.\ref{e:sort-code}.

\paragraph{How to speed up the slowsort (Ex.~\ref{e:slow-sort})}
There is a paper about that: \cite{fischer-purely}.

\end{document}